\documentclass{llncs}

\usepackage{mathtools,amssymb,amsfonts}    
\usepackage{nicefrac}
\usepackage{stmaryrd}                      
\usepackage[textsize=small, textwidth=12ex]{todonotes}                     
\usepackage[inline]{enumitem}                      
\usepackage{xspace}                        
\usepackage{tikz}
\usetikzlibrary{arrows,decorations.pathmorphing,positioning,fit,trees,shapes,shadows,automata,calc}
\usepackage{pgfplots}
\usepackage{leftidx}
\usepackage{framed}
\usepackage{needspace}
\usepackage{graphicx}
\usepackage{wrapfig}
\usepackage{todonotes}
\usepackage{fancybox}
\usepackage{multirow}
\usepackage{multicol}
\usepackage{subfigure}
\usepackage{hyperref}

\definecolor{orange0}{RGB}{255,255,0}
\definecolor{orange1}{RGB}{255,220,0}
\definecolor{orange2}{RGB}{255,180,0}
\definecolor{orange3}{RGB}{255,140,0}
\definecolor{orange4}{RGB}{255,90,0}
\definecolor{orange5}{RGB}{255,0,0}

\definecolor{blue0}{RGB}{140,170,255}
\definecolor{blue1}{RGB}{120,150.6,255}
\definecolor{blue2}{RGB}{100,130.6,255}
\definecolor{blue3}{RGB}{80,110,255}
\definecolor{blue4}{RGB}{60,90,255}

\definecolor{codegreen}{rgb}{0,0.6,0}
\definecolor{codegray}{rgb}{0.5,0.5,0.5}
\definecolor{codepurple}{rgb}{0.58,0,0.82}
\definecolor{backcolour}{rgb}{0.95,0.95,0.92}
\definecolor{gray}{rgb}{0.4,0.4,0.4}
\definecolor{darkblue}{rgb}{0.0,0.0,0.6}
\definecolor{cyan}{rgb}{0.0,0.6,0.6}
\definecolor{whiteorange}{RGB}{255,224,171}

\usepackage{listings}          
\lstdefinestyle{pseudocode}{
	basicstyle=\footnotesize\ttfamily,
	backgroundcolor=\color{white},   
	commentstyle=\color{codegreen},
	keywordstyle=\color{magenta},
	numberstyle=\tiny\color{codegray},
	stringstyle=\color{codepurple},
	breakatwhitespace=false,         
	breaklines=true,
	captionpos=b,                    
	keepspaces=true,
	numbers=left,
	numbersep=5pt,                  
	showspaces=false,                
	showstringspaces=false,
	showtabs=false,                  
	tabsize=4
}

\makeatletter
\newenvironment{CenteredBox}{%
	\begin{Sbox}}{
	\end{Sbox}\centerline{\parbox{\wd\@Sbox}{\TheSbox}}}
\makeatother

\usepackage{ushort}
\newcommand{\ie}{i.e.\@\xspace}
\newcommand{\wrt}{w.r.t.\@\xspace}
\newcommand{\eg}{e.g.\@\xspace}

\newcommand{\prophesy}{\textrm{PROPhESY}\xspace}
\newcommand{\prism}{\textrm{PRISM}\xspace}
\newcommand{\storm}{\textrm{Storm}\xspace}
\newcommand{\tool}[1]{\textrm{#1}\xspace}

\newcommand{\dtmc}{\mathcal{D}}

\newcommand{\p}{\ensuremath{\mathbb{P}}}
\newcommand{\pr}{\ensuremath{\mathrm{Pr}}}
\newcommand{\reachPr}[2]{\ensuremath{\pr^{#1}(\finally #2)}}
\newcommand{\reachPrT}[1][]{\ensuremath{\reachPr{#1}{T}}}

\newcommand{\reachPropSymbol}{\varphi}
\newcommand{\ereachPropSymbol}{\psi}

\newcommand{\rew}{\ensuremath{\mathbb{E}}}
\newcommand{\er}{\ensuremath{\mathrm{ER}}}
\newcommand{\expRew}[2]{\ensuremath{\er^{#1}(\finally #2)}}
\newcommand{\expRewT}[1][]{\ensuremath{\expRew{#1}{T}}}

\newcommand{\rewFunction}{\ensuremath{\mathrm{rew}}}
\newcommand{\finally}{\lozenge}

\newcommand{\creachPr}[3]{\ensuremath{\pr^{#1}(\finally #2 \mathrel{|} \neg \finally #3)}}
\newcommand{\creachPrT}[1][]{\ensuremath{\creachPr{#1}{T}{U}}}
\newcommand{\creachProp}[3]{\ensuremath{\p_{\leq #1}(\finally #2 \mathrel{|} \neg \finally #3)}}
\newcommand{\creachProplT}{\ensuremath{\creachProp{\lambda}{T}{U}}}

\newcommand{\cexpRew}[3]{\ensuremath{\er^{#1}(\finally #2 \mathrel{|} \neg \finally #3)}}
\newcommand{\cexpRewT}[1][]{\ensuremath{\cexpRew{#1}{T}{U}}}

\newcommand{\cexpRewProp}[3]{\ensuremath{\rew_{\leq #1}(\finally #2 \mathrel{|} \neg \finally #3)}}
\newcommand{\cexpRewPropkT}{\ensuremath{\cexpRewProp{\kappa}{T}{U}}}

\newcommand{\nj}[1]{\todo[inline,color=yellow!50]{\color{black}NJ: #1}\color{black}}

\newcommand{\R}{\mathbb{R}}
\newcommand{\Q}{\mathbb{Q}}

\newcommand{\Ireal}{[0,\, 1]\subseteq\mathbb{R}}  
\newcommand{\Ex}{\ensuremath{\mathbb{E}}\xspace}        

\newcommand{\Distr}{\mathit{Distr}}

\newcommand{\distDom}{X}

\newcommand{\distFunc}{\mu}
\newcommand{\distDomElem}{x}

\newcommand{\true}{\ensuremath{\mathsf{true}}\xspace}

\newcommand{\pGCL}{\textnormal{\textsf{pGCL}}\xspace}

\newcommand{\cpGCL}{\textnormal{\textsf{cpGCL}}\xspace}

\newcommand{\Cmd}{\ensuremath{\mathcal{P}}\xspace}        
\newcommand{\Var}{\ensuremath{\mathcal{V}}\xspace}        
\newcommand{\Paramvar}{\ensuremath{{V}}\xspace}        

\newcommand{\State}{\ensuremath{\mathbb{S}}\xspace}      

\newcommand{\Skip}{{\tt skip}\xspace}
\newcommand{\Abort}{{\tt abort}\xspace}
\newcommand{\Ass}[2]{#1 \coloneqq #2}
\newcommand{\Observe}{\textnormal{\texttt{observe}}\xspace}
\newcommand{\PChoice}[3]{\{#1\} \, \left[#2 \right] \, \{#3\}}
\newcommand{\NDChoice}[2]{\{#1\} \, \Box \, \{#2\}}
\newcommand{\If}{{\tt if}\xspace}
\newcommand{\Then}{{\tt then}\xspace}
\newcommand{\Else}{{\tt else}\xspace}
\newcommand{\While}{{\tt while}\xspace}

\newcommand{\Ite}{{\tt ite}\xspace}
\newcommand{\Cond}[3]{{\Ite} \, (#1) \allowbreak\, \{#2\} \allowbreak\, \{#3\}}
\newcommand{\WhileDo}[2]{{\While} \,\allowbreak (#1) \,\allowbreak \{#2\}}

\newcommand{\subst}[2]{[#1 / #2]}

\newcommand{\sinit}{s_{\mathit{I}}} 
\newcommand{\mdp}{\mathcal{M}}
\newcommand{\MdpInit}[1][]{\ensuremath{\mdp{#1}=(S{#1},\,\sinit{#1},\,\Act,\,\probmdp{#1})}}
\newcommand{\probmdp}{\mathcal{P}}
\newcommand{\functions}[1][]{\ensuremath{\mathbb{Q}_{#1}}}
\newcommand{\poly}[1][]{\ensuremath{\mathbb{Q}[#1]}}

\newcommand{\pol}{\ensuremath{g}}

\newcommand{\pdtmc}{\ensuremath{\mathcal{P}}}

\renewcommand{\Pr}{\ensuremath{\textnormal{Pr}}}

\newcommand{\sched}{\ensuremath{\mathfrak{S}}}
\newcommand{\Sched}{\ensuremath{\mathit{Sched}}}

\newcommand{\Rmdp}[3]{\ensuremath{\mdp_{#1}^{#2}\llbracket #3\rrbracket}}
\newcommand{\Act}{\ensuremath{\mathit{Act}}}
\newcommand{\act}{\ensuremath{\alpha}}
\newcommand{\pmdp}{\ensuremath{\mathcal{P}}}

\newcommand{\bad}{\ensuremath{\lightning}}
\newcommand{\undesired}{\ensuremath{\langle \bad \rangle}\xspace}
\newcommand{\sinklabel}{\mathpzc{sink}\xspace}
\newcommand{\sink}{\ensuremath{\langle \sinklabel \rangle}\xspace}

\newcommand{\term}{\ensuremath{{\downarrow}}}

\newcommand{\exit}{\ensuremath{{\downarrow}}}

\DeclareMathAlphabet{\mathpzc}{OT1}{pzc}{m}{it}
\def\presuper#1#2%
  {\mathop{}%
   \mathopen{\vphantom{#2}}^{#1}%
   \kern-\scriptspace%
   #2}

\allowdisplaybreaks

\title {Bounded Model Checking for\\ Probabilistic Programs\thanks{This work has been partly funded by the awards AFRL \# FA9453-15-1-0317, ARO \# W911NF-15-1-0592 and ONR \# N00014-15-IP-00052 and is supported by the Excellence Initiative of the German federal and state government.}}

\author{Nils Jansen\inst{2} \and Christian Dehnert\inst{1}\and Benjamin Lucien Kaminski\inst{1} \and \\ Joost-Pieter Katoen\inst{1} \and Lukas Westhofen\inst{1}}

\institute{RWTH Aachen University, Germany \and University of Texas at Austin, USA}

\begin{document}
\maketitle

\begin{abstract}
	In this paper we investigate the applicability of standard model checking approaches to verifying properties in probabilistic programming. 
As the operational model for a standard probabilistic program is a potentially infinite parametric Markov decision process, no direct adaption of existing techniques is possible. 
Therefore, we propose an on--the--fly approach where the operational model is successively created and verified via a step--wise execution of the program. 
This approach enables to take key features of many probabilistic programs into account: 
nondeterminism and conditioning. 
We discuss the restrictions and demonstrate the scalability on several benchmarks.
\end{abstract}

\section{Introduction}
Probabilistic programs are imperative programs, written in languages like \texttt{C}, \texttt{Scala}, \texttt{Prolog}, or \texttt{ML}, with two added constructs: (1) the ability to draw values at random from probability distributions, and (2) the ability to condition values of variables in a program
through observations. 
In the past years, such programming languages became very popular due to their wide applicability for several different research areas~\cite{DBLP:conf/icse/GordonHNR14}: 
Probabilistic programming is at the heart of \emph{machine learning} for describing distribution functions; \emph{Bayesian inference} is pivotal in their analysis.
They are central in \emph{security} for describing cryptographic constructions (such as randomized encryption) and security experiments. 
In addition, probabilistic programs are an active research topic in \emph{quantitative information flow}. 
Moreover, \emph{quantum programs} are inherently probabilistic due to the random outcomes of quantum measurements.
All in all, the simple and intuitive syntax of probabilistic programs makes these different research areas accessible to a broad audience. 

However, although these programs typically consist of a few lines of code, they  are often hard to understand and analyze; bugs, for instance \emph{non--termination} of a program, can easily occur. It seems of utmost importance to be able to automatically prove properties like \emph{``Is the probability for termination of the program at least 90\%''} or \emph{``Is the expected value of a certain program variable at least 5 after successful termination?''}. 
Approaches based on the simulation of a program to show properties or infer probabilities have been made in the past~\cite{DBLP:conf/pldi/SankaranarayananCG13,DBLP:conf/sigsoft/ClaretRNGB13}. However, to the best of our knowledge there is no work which exploits well-established \emph{model checking algorithms} for probabilistic systems such as Markov decision processes (MDP) or Markov chains (MCs), as already argued to be an interesting avenue for the future in~\cite{DBLP:conf/icse/GordonHNR14}.

As the operational semantics for a probabilistic program can be expressed as a (possible infinite) MDP~\cite{DBLP:journals/pe/GretzKM14}, it seems worthwhile to investigate the opportunities there. However, probabilistic model checkers like \prism~\cite{KNP11}, \tool{iscasMc}~\cite{DBLP:conf/fm/HahnLSTZ14}, or \tool{MRMC}~\cite{mrmc} offer efficient methods only for \emph{finite models}. 

We make use of the simple fact that for a finite unrolling of a program the corresponding operational MDP is also finite. Starting from a profound understanding of the (intricate) probabilistic program semantics---including features
such as observations, unbounded (and hence possibly diverging) loops, and nondeterminism---we show that with each unrolling of the program both conditional reachability probabilities and conditional expected values of program variables increase monotonically. This gives rise to a \emph{bounded model-checking approach} for verifying probabilistic programs. This enables for a user to write a program and automatically verify it against a desired property without further knowledge of the programs semantics.

We extend this methodology to the even more complicated case of \emph{parametric probabilistic programs}, where probabilities are given by functions over parameters. At each iteration of the bounded model checking procedure, parameter valuations violating certain properties are guaranteed to induce violation at each further iteration. 

We demonstrate the applicability of our approach using five well-known benchmarks from the literature. Using efficient model building and verification methods, our prototype is able to prove properties where either the state space of the operational model is infinite or consists of millions of states.

\paragraph{Related Work.} Besides the tools employing probabilistic model checking as listed above, one should mention the approach in~\cite{Kat11}, where \emph{finite abstractions} of the operational semantics of a program were verified. However, this was defined for programs without parametric probabilities or observe statements. In~\cite{DBLP:journals/siamcomp/SharirPH84}, verification on partial operational semantics is theoretically discussed for termination probabilities.

The paper is organized as follows: In Section~\ref{sec:preliminaries}, we introduce the probabilistic models we use, the probabilistic programming language, and the structured operational semantics (SOS) rules to construct an operational (parametric) MDP. Section~\ref{sec:bmc} first introduces formal concepts needed for the finite unrollings of the program, then shows how expectations and probabilities grow monotonically, and finally explains how this is utilized for bounded model checking. In Section~\ref{sec:evaluation}, an extensive description of used benchmarks, properties and experiments is given before the paper concludes with Section~\ref{sec:conclusion}.

\section{Preliminaries}\label{sec:preliminaries}
\subsection{Distributions and Polynomials}

A \emph{probability distribution} over a finite or countably infinite set $\distDom$ is a function $\distFunc\colon\distDom\rightarrow\Ireal$ with $\sum_{\distDomElem\in\distDom}\distFunc(\distDomElem)=1$. 
The set of all distributions on $\distDom$ is denoted by $\Distr(\distDom)$.
Let $\Paramvar$ be a finite set of \emph{parameters} over $\R$.
A \emph{valuation} for $\Paramvar$ is a function $u \colon \Paramvar \to \R$.
Let $\poly[\Paramvar]$ denote the set of multivariate \emph{polynomials} with rational coefficients and $\functions[\Paramvar]$ the set of \emph{rational functions} (fractions of polynomials) over $\Paramvar$. For $\pol\in \poly[\Paramvar]$ or $\pol\in\functions[\Paramvar]$, let $\pol[u]$ denote the evaluation of $\pol$ at $u$. We write $\pol=0$ if $\pol$ can be reduced to $0$, and $\pol\not=0$ otherwise.


\subsection{Probabilistic Models}\label{sec:probmodels}

First, we introduce parametric probabilistic models which can be seen as transition systems where the transitions are labelled with polynomials in $\poly[\Paramvar]$.
\begin{definition}[pMDP and pMC]\label{def:pmdp}
A \emph{parametric Markov decision process (pMDP)} is a tuple $\MdpInit$ with a countable set $S$ of states, an initial state $\sinit \in S$, a finite set $\Act$ of actions, and a transition function $\probmdp \colon S \times \Act \times S \rightarrow \poly[\Paramvar]$ satisfying for all $s\in S\colon
\Act(s) \neq \emptyset$, where $\Paramvar$ is a finite set  of parameters over $\R$ and $\Act(s) = \{\act \in \Act \mid \exists s'\in S.\,\probmdp(s,\,\act,\,s') \neq 0\}$.
If for all $s\in S$ it holds that $|\Act(s)| = 1$, $\mdp$ is called a \emph{parametric discrete-time Markov chain (pMC)}, denoted by $\dtmc$.
\end{definition}
At each state, an action is chosen \emph{nondeterministically}, then the successor states are determined \emph{probabilistically} as defined by the transition function.
$\Act(s)$ is the set of \emph{enabled} actions at state $s$.
As $\Act(s)$ is non-empty for all $s \in S$, there are no deadlock states. For pMCs there is only one single action per state and we write the transition probability function as $\probmdp\colon S\times S\rightarrow\poly[\Paramvar]$, omitting that action.
\emph{Rewards} are defined using a \emph{reward function} $\rewFunction \colon S \rightarrow \R$ which assigns rewards to states of the model.
Intuitively, the reward $\rewFunction(s)$ is earned upon \emph{leaving} the state $s$. 

\paragraph{Schedulers.}

The nondeterministic choices of actions in pMDPs can be resolved using \emph{schedulers}\footnote{Also referred to as adversaries, strategies, or policies.}.
In our setting it suffices to consider memoryless deterministic schedulers~\cite{Var85}. 
For more general definitions we refer to~\cite{BK08}.
\begin{definition}{\bf (Scheduler)}\label{def:scheduler}
	A \emph{scheduler} for pMDP $\MdpInit$ is a function $\sched\colon S\rightarrow\Act$ with $\sched(s)\in \Act(s)$ for all $s\in S$.  
\end{definition}
Let $\Sched^\mdp$ denote the set of all schedulers for $\mdp$.
Applying a scheduler to a pMDP yields an \emph{induced parametric Markov chain}, as all nondeterminism is resolved, \ie, the transition probabilities are obtained \wrt the choice of actions.
\begin{definition}{\bf (Induced pMC)}\label{def:induced_dtmc} 
	Given a pMDP $\MdpInit$, the \emph{pMC induced by $\sched\in\Sched^\mdp$} is given by $\mdp^\sched=(S,\, \sinit,\, \Act,\, \probmdp^\sched)$, where
	\begin{align*}
		\probmdp^\sched(s,\,s')= \probmdp(s,\, \sched(s),\, s'), \quad \mbox{for all } s,s'\in S~.
	\end{align*} 
\end{definition}
%
%

\paragraph{Valuations.}

Applying a \emph{valuation} $u$ to a pMDP $\mdp$, denoted $\mdp[u]$, replaces each polynomial $\pol$ in $\mdp$ by $\pol[u]$.
We call $\mdp[u]$ the \emph{instantiation} of $\mdp$ at $u$.
A valuation $u$ is \emph{well-defined} for $\mdp$ if the replacement yields \emph{probability distributions} at all states; the resulting model $\mdp[u]$ is a Markov decision process (MDP) or, in absence of nondeterminism, a Markov chain (MC).

\paragraph{Properties.}

For our purpose we consider \emph{conditional reachability properties} and \emph{conditional expected reward properties} in MCs.
For more detailed definitions we refer to~\cite[Ch.\ 10]{BK08}.
Given an MC $\dtmc$ with state space $S$ and initial state $\sinit$, let $\pr^{\dtmc}(\neg \finally U)$ denote the probability \emph{not} to reach a set of undesired states $U$ from the initial state $\sinit$ within $\dtmc$.
Furthermore, let $\creachPr{\dtmc}{T}{U}$ denote the conditional probability to reach a set of target states $T \subseteq S$ from the initial state $\sinit$ within $\dtmc$, given that no state in the set $U$ is reached.
We use the standard probability measure on infinite paths through an MC.
For threshold $\lambda\in \Ireal$, the reachability property, asserting that a target state is to be reached with conditional probability at most $\lambda$, is denoted $\reachPropSymbol = \creachProplT$.
The property is satisfied by $\dtmc$, written $\dtmc \models \reachPropSymbol$, iff $\creachPrT[\dtmc]\leq\lambda$.
This is analogous for comparisons like $<$, $>$, and $\geq$.

The reward of a path through an MC $\dtmc$ until $T$ is the sum of the rewards of the states visited along on the path before reaching $T$.
The expected reward of a finite path is given by its probability times its reward.
Given $\reachPrT[\dtmc] = 1$, the conditional expected reward of reaching $T \subseteq S$, given that no state in set $U \subseteq S$ is reached, denoted $\cexpRewT[\dtmc]$, is the expected reward of all paths accumulated until hitting $T$ while not visiting a state in $U$ in between divided by the probability of not reaching a state in $U$ (\ie, divided by $\pr^{\dtmc}(\neg \finally U)$).
An expected reward property is given by $\ereachPropSymbol = \cexpRewPropkT$ with threshold $\kappa \in \R_{\geq 0}$.
The property is satisfied by $\dtmc$, written $\dtmc \models \ereachPropSymbol$, iff $\cexpRewT[\dtmc]\leq\kappa$.
Again, this is analogous for comparisons like $<$, $>$, and $\geq$.
%
For details about conditional probabilities and expected rewards see~\cite{DBLP:conf/tacas/BaierKKM14}. 

Reachability probabilities and expected rewards for MDPs are defined on induced MCs for specific schedulers. 
We take here the conservative view that a property for an MDP has to hold for \emph{all possible schedulers}. 

\paragraph{Parameter Synthesis.}
For pMCs, one is interested in \emph{synthesizing} well-defined valuations that induce satisfaction or violation of the given specifications~\cite{dehnert-et-al-cav-2015}. In detail, for a pMC $\dtmc$, a rational function $\pol\in\functions[\Paramvar]$ is computed which---when instantiated by a well-defined valuation $u$ for $\dtmc$---evaluates to the actual reachability probability or expected reward for $\dtmc$, \ie, $\pol[u]=\reachPr{\dtmc[u]}{T}$ or $\pol[u]=\expRewT[{\dtmc[u]}]$. For pMDPs, schedulers inducing \emph{maximal} or \emph{minimal} probability or expected reward have to be considered~\cite{quatmann-et-al-techreport-2016}.

\subsection{Conditional Probabilistic Guarded Command Language}

We first present a programming language which is an extension of Dijkstra's guarded command language~\cite{Dijkstra} with a binary probabilistic choice operator, yielding the 
\emph{probabilistic guarded command language} (\pGCL)~\cite{McIver:2004}. In~\cite{jansen-et-al-mfps-2015}, \pGCL was endowed with \emph{observe statements}, giving rise to conditioning. The syntax of this \emph{conditional probabilistic guarded command language} (\cpGCL) is given by
\begin{align*}
\Cmd ~\Coloneqq~  &\Skip \mid \Abort \mid \Ass x E \mid \Cmd;\Cmd \mid \If\ G\ \Then\ {\Cmd }\ \Else\ {\Cmd } \\ 
   &{}~\mid  \, \PChoice {\Cmd} {\pol} {\Cmd} \mid
\NDChoice {\Cmd}{\Cmd} \mid \WhileDo G \Cmd \mid \Observe~(G)
\end{align*}
%
%
 Here, $x$ belongs to the set of \emph{program variables} $\Var$; $E$ is an arithmetical expression over $\Var$; $G$ is a \emph{Boolean expression} over arithmetical expressions over $\Var$. The \emph{probability} is given by a polynomial $\pol \in \poly[\Paramvar]$.
 Most of the \cpGCL instructions are self--explanatory; we elaborate only on the following: For \cpGCL-programs $P$ and $Q$, $\PChoice{P}{\pol}{Q}$ is a \emph{probabilistic choice} where $P$ is executed with probability $\pol$ and $Q$ with probability $1{-}\pol$; analogously, $\NDChoice{P}{Q}$ is a \emph{nondeterministic choice} between $P$ and $Q$;  \Abort is syntactic sugar for the diverging program $\While\ (\true)\ \{\Skip\}$. The statement $\Observe~(G)$ for the Boolean expression $G$ \emph{blocks} all program executions violating $G$ and induces a \emph{rescaling} of probability of the
remaining execution traces so that they sum up to one. 
For a \cpGCL-program $P$, the set of \emph{program states} is given by $\State = \{\sigma ~|~ \sigma \colon \Var \to \Q\}$, 
\ie, the set of all  variable valuations. We assume all variables to be assigned zero prior to execution or at the start of the program. This initial variable valuation $\sigma_I\in\State$ with $\forall x \in \Var.\, \sigma_I(x) = 0$ is called the \emph{initial state} of the program. 
\begin{example}\label{ex:simple}
	Consider the following \cpGCL-program with variables $x$ and $c$:

			\begin{CenteredBox}
\begin{lstlisting}[language=C,linewidth=9cm]
while (c = 0) {
       { x := x + 1 } [0.5] { c := 1 }
};
observe "x is odd"
\end{lstlisting}
			\end{CenteredBox}

\noindent While $c$ is $0$, the loop body is iterated: With probability $\nicefrac 1 2$ either $x$ is incremented by one or $c$ is set to one. After leaving the loop, the event that the valuation of $x$ is odd is \emph{observed}, which means that all program executions where $x$ is even are blocked. 
	Properties of interest for this program would, \eg, concern the termination probability, or the expected value of $x$ after termination.
\hfill$\triangle$
\end{example}

\subsection{Operational Semantics for Probabilistic Programs}\label{sec:operational}
We now introduce an operational semantics for \cpGCL-programs which is given by an MDP as in Definition~\ref{def:pmdp}.
The structure of such an operational MDP is schematically depicted below.
	\begin{center}
		\scalebox{0.9}{\begin{tikzpicture}[->,>=stealth',shorten >=1pt,node distance=2.5cm,semithick,minimum size=1cm]
\tikzstyle{every state}=[draw=none]
  \draw[white, use as bounding box] (-1.2,-1.8) rectangle (6.5,1.5);
   \node [state, initial, initial text=] (init) {$\langle P,\, \sigma_I \rangle$};  
   \node [cloud, draw=black,cloud puffs=15, cloud puff arc= 150,
        minimum width=1.5cm, minimum height=.75cm, aspect=1] (exit) [right of=init] {$\exit$};
   \node [state] (bad) [above=0.3cm of exit] {$\undesired$};
   \node [state] (sink) [right of=exit] {$\sink$};
   \node [cloud, draw=black,cloud puffs=15, cloud puff arc= 150,
        minimum width=1.5cm, minimum height=.75cm, aspect=1] (diverge) [below=0.5 cm of exit] {$\phantom{\exit}$};

    \node [] (divergetext) [below=-0.825 cm of diverge] {\small$\mathpzc{diverge}$};

   \node [state] (haken1) at (2.2, .1) {\scriptsize $\exit$};
   \node [state] (haken2) at (2.25, -.1) {\tiny $\exit$};
    \node [state] (haken3) at (2.7, -.1) {\scriptsize $\exit$};
   \node [state] (haken4) at (2.75, .1) {\tiny $\exit$};
   \node [state] (haken5) at (2.95, .0) {\tiny $\exit$};

  \path [] 
      (init) edge [decorate,decoration={snake, post length=2mm}] (exit)
      (init) edge [decorate,decoration={snake, post length=2mm}] (bad)
      (init) edge [decorate,decoration={snake, post length=2mm}] (diverge)
      (exit) edge [] (sink)
      (bad) edge [] (sink)
      (sink) edge [loop right] (sink)
      (diverge) edge [loop right,decorate,decoration={snake, post length=2mm}] (diverge)
  ;
\end{tikzpicture}
}
	\end{center}
Squiggly arrows indicate reaching certain states via possibly multiple paths and states; the clouds indicate that there might be several states of the particular kind.
$\langle P,\, \sigma_I \rangle$ marks the initial state of the program $P$. 
In general the states of the operational MDP are of the form $\langle P',\, \sigma' \rangle$ where $P'$ is the program that is left to be executed and $\sigma'$ is the current variable valuation.

All runs of the program (paths through the MDP) are either \emph{terminating} and eventually end up in the \sink state, or are \emph{diverging} (thus they never reach \sink).
Diverging runs occur due to non--terminating computations.
A terminating run has either terminated successfully, \ie, it passes a \exit--state, or it has terminated due to a \emph{violation of an observation}, \ie, it passes the $\undesired$--state.
Sets of runs that eventually reach \undesired, or \sink, or diverge are pairwise disjoint. 

\noindent The \exit--labelled states are the \emph{only ones} with positive reward, which is due to the fact that we want to capture probabilities of events (respectively expected values of random variables) occurring at \emph{successful termination} of the program.

The random variables of interest are $\Ex = \{f ~|~ f\colon \State \to \R_{\geq 0}\}$.
Such random variables are referred to as post--expectations~\cite{McIver:2004}. 
Formally, we have:
\begin{definition}[Operational Semantics of Programs]
The \emph{operational semantics} of a \cpGCL program $P$ with respect to a post--expectation $f \in \Ex$ is the MDP $\Rmdp{}{f}{P} = (S$, $\langle P,\,\sigma_I\rangle,\, \Act,\, \pdtmc)$ together with a reward function $\rewFunction$, where
\begin{itemize}
\item$S = \big\{\langle Q,\, \sigma \rangle,\langle
\term,\, \sigma \rangle ~\big|~ Q \textnormal{ is a \cpGCL program},\, \sigma \in \State\big\} \cup \{\undesired,\, \sink\}$ is the countable set of states,
\item $\langle P,\,\sigma_I\rangle\in S$ is the initial state,
\item $\Act=\{\mathit{left},\,\mathit{right},\, \mathit{none}\}$ is the set of actions, and
\item $\probmdp$ is the smallest relation defined by the SOS rules given in Figure \ref{fig:sos-rules}.
\end{itemize}
The reward function is $\rewFunction(s) = f(\sigma)$ if $s = \langle\term,\, \sigma\rangle$, and $\rewFunction(s) = 0$, otherwise.
\end{definition}
\begin{figure}[t]
\scriptsize
\begin{align*}
&(\textbf{terminal})\,\frac{\vphantom{\langle}}{\langle {\downarrow},\, \sigma \rangle ~\longrightarrow~ \sink} \quad
(\textbf{skip})\,\frac{\vphantom{\langle}}{\langle \Skip,\, \sigma \rangle ~\longrightarrow~ \langle {\downarrow},\, \sigma \rangle} \quad
(\textbf{abort})\,\frac{\vphantom{\langle}}{\langle \Abort,\, \sigma \rangle ~\longrightarrow~ \langle \Abort,\, \sigma \rangle}\\
&(\textbf{undesired})\,\frac{\vphantom{\langle}}{\undesired ~\longrightarrow~ \sink}\qquad\qquad\qquad\qquad\qquad~~~~
(\textbf{assign})\,\frac{\vphantom{\langle}}{\langle \Ass x E,\, \sigma \rangle ~\longrightarrow~ \langle {\downarrow},\, \sigma[x \leftarrow \llbracket E \rrbracket_\sigma] \rangle}\\
&(\textbf{observe1})\,\frac{\sigma \models G}{\langle \Observe\, G,\, \sigma \rangle ~\longrightarrow~ \langle {\downarrow} ,\, \sigma \rangle}\qquad\qquad\qquad\qquad\qquad\,
(\textbf{observe2})\,\frac{\sigma \not\models G}{\langle \Observe\, G,\, \sigma \rangle ~\longrightarrow~ \undesired}\\
&(\textbf{concatenate1})\,\frac{}{\langle\downarrow;{Q},\, \sigma \rangle ~\longrightarrow~ \langle Q,\, \sigma \rangle}\qquad\qquad\qquad\qquad\qquad\;
(\textbf{concatenate2})\,\frac{\langle P,\, \sigma \rangle ~\longrightarrow~ \undesired}{\langle {P};{Q},\, \sigma \rangle ~\longrightarrow~ \undesired}\\
&(\textbf{concatenate3})\,\frac{\langle P,\, \sigma \rangle ~\longrightarrow~ \mu}{\langle {P};{Q},\, \sigma \rangle ~\longrightarrow~ \nu}, 
\textnormal{where } \forall P'.\,\nu(\langle {P'};{Q}, \sigma'\rangle) := \mu(\langle P',\, \sigma'\rangle)\\
&(\textbf{if1})\,\frac{\sigma \models G}{\langle \Cond G P Q,\, \sigma \rangle ~\longrightarrow~ \langle P,\, \sigma \rangle}\qquad\qquad\qquad\qquad~~\;
(\textbf{if2})\,\frac{\sigma \not\models G}{\langle \Cond G P Q,\, \sigma \rangle ~\longrightarrow~ \langle Q,\, \sigma \rangle}\\
&(\textbf{while1})\,\frac{\sigma \models G}{\langle \WhileDo G P,\, \sigma \rangle ~\longrightarrow~ \langle {P};{\WhileDo G P},\, \sigma \rangle}~~\,
(\textbf{while2})\,\frac{\sigma \not\models G}{\langle \WhileDo G P,\, \sigma \rangle ~\longrightarrow~ \langle {\downarrow},\, \sigma \rangle}\\
&(\textbf{prob})\,\frac{\vphantom{\langle}}{\langle \PChoice P p Q,\, \sigma \rangle ~\longrightarrow~ \nu}, \textnormal{where } \nu(\langle P,\, \sigma\rangle) := p,\, \nu(\langle Q,\, \sigma\rangle) := 1 - p\\
&(\textbf{nondet1})\,\frac{\vphantom{\langle}}{\langle \NDChoice P Q ,\, \sigma \rangle ~ \xrightarrow{~\mathit{left}~} ~ \langle P,\, \sigma \rangle}\qquad\qquad\qquad
(\textbf{nondet2})\frac{\vphantom{\langle}}{\langle \NDChoice P Q ,\, \sigma \rangle ~\xrightarrow{~\mathit{right}~}~ \langle Q,\, \sigma \rangle}
\end{align*}\normalsize
\caption{
SOS rules for constructing the operational MDP of a \cpGCL program.
We use $s \longrightarrow t$ to indicate $\probmdp(s,\, \mathit{none},\, t) = 1$, $s \longrightarrow \mu$ for $\mu \in \Distr(S)$ to indicate $\forall t \in S\colon \probmdp(s,\, \mathit{none},\, t) = \mu(t)$, $s \xrightarrow{~\mathit{left}~} t$ to indicate $\probmdp(s,\, \mathit{left},\, t) = 1$, and $s \xrightarrow{~\mathit{right}~} t$ to indicate $\probmdp(s,\, \mathit{right},\, t) = 1$.
}
\label{fig:sos-rules}
\end{figure}
A state of the form $\langle {\downarrow},\, \sigma \rangle$ indicates successful termination, \ie, no commands are left to be executed.
These terminal states and the \undesired--state go to the \sink state. 
\Skip without context terminates successfully.
\Abort self--loops, \ie, diverges. 
$\Ass x E$ alters the variable valuation according to the assignment then terminates successfully. 
For the concatenation, $\langle\term;{Q},\, \sigma \rangle$ indicates successful termination of the first program, so the execution continues with $\langle{Q},\, \sigma \rangle$.
If for $P;\,Q$ the execution of $P$ leads to $\undesired$, $P;\,Q$ does so, too.
Otherwise, for $\langle P,\sigma \rangle{\longrightarrow}\mu$, $\mu$ is lifted such that $Q$ is concatenated to the support of $\mu$. For more details on the operational semantics we refer to~\cite{DBLP:journals/pe/GretzKM14}.

If for the conditional choice $\sigma\models G$ holds, $P$ is executed, otherwise $Q$.
The case for $\While$ is similar.
For the probabilistic choice, a distribution $\nu$ is created according to probability $p$. 
For $\NDChoice P Q$, we call $P$ the $\mathit{left}$ choice and $Q$ the $\mathit{right}$ choice for actions $\mathit{left}, \mathit{right}\in\Act$.
For the \Observe statement, if $\sigma\models G$ then \Observe acts like $\Skip$. 
Otherwise, the execution leads directly to \undesired indicating a violation of the \Observe statement.
\begin{figure}[t]
\begin{center}
\scalebox{0.8}{\begin{tikzpicture}[->,>=stealth',shorten >=1pt,semithick,minimum size=.5cm]
\tikzstyle{every state}=[draw=none]
   \node [] (init) at (0,0) {$\langle P,\,\sigma_I \rangle$};  
    \node [] (s1) [on grid, below=1.1 cm of init] {$\langle P_1;P,\,\sigma_I\rangle$};   
    \node [] (s11) [on grid, left=4 cm of s1] {$\langle P_3;\,P,\,\sigma_I\rangle$};   
    \node [] (s12) [on grid, right=4 cm of s1] {$\langle P_4;\,P,\,\sigma_I\rangle$};   
	\node [] (s111) [on grid, below=1.1 cm of s11] {$\langle \term;\,P,\,\sigma_I[x/1]\rangle$};   
    \node [] (s121) [on grid, below=1.1 cm of s12] {$\langle \term;\,P,\,\sigma_I[c/1]\rangle$};   
    \node [] (s1111) [on grid, below=1.1 cm of s111] {$\langle P,\,\sigma_I[x/1]\rangle$};   
    \node [] (s1211) [on grid, below=1.1 cm of s121] {$\langle P,\,\sigma_I[c/1]\rangle$};   
    \node [] (s1211b) [on grid, below=1.1 cm of s1211] {$\langle \term;\,P_2,\,\sigma_I[c/1]\rangle$};   
    \node [] (s1211bb) [on grid, below=1.1 cm of s1211b] {$\langle P_2,\,\sigma_I[c/1]\rangle$}; 
    \node [] (s11111) [on grid, left=4 cm of s1111] {$\langle P_3;P,\,\sigma_I[x/1]\rangle$};   
    \node [] (s11112) [on grid, right=4 cm of s1111] {$\langle P_4;P,\,\sigma_I[x/1]\rangle$};   
    \node [] (s111111) [on grid, below=1.1 cm of s11111] {$\langle \term;P,\,\sigma_I[x/2]\rangle$};   
    \node [] (s111121) [on grid, below=1.1 cm of s11112] {$\langle \term;P,\,\sigma_I[x/1,c/1]\rangle$};   
    \node [] (s1111111) [on grid, below=1.1 cm of s111111] {$\langle P,\,\sigma_I[x/2]\rangle$};   
    \node [] (s1111211) [on grid, below=1.1 cm of s111121] {$\langle P,\,\sigma_I[x/1,c/1]\rangle$};
    \node [] (s1111211b) [on grid, below=1.1 cm of s1111211] {$\langle \term;\,P_2,\,\sigma_I[x/1,c/1]\rangle$};
    \node [] (s1111211bb) [on grid, below=1.1 cm of s1111211b] {$\langle P_2,\,\sigma_I[x/1,c/1]\rangle$};
    \node [label={[yshift=-0.5cm] 170:\large\boxed{$1$}}] (s11112111) [on grid, below=1.1 cm of s1111211bb] {$\langle \term,\,\sigma_I[x/1,c/1]\rangle$};
%

  \node [] (undesired) [on grid, below=1.1 cm of s1211bb] {$\undesired$};  
  \node [] (sink) [on grid, below=2.2 cm of undesired] {$\sink$};  
  \node [] (dots) [on grid, below=.5 cm of s1111111] {$\bf\vdots$};  
	
\path
      (init) edge [] node [left, near start] {} (s1)
      (s1) edge [] node [above, near start] {\scriptsize{$\frac{1}{2}$}} (s11)
      (s1) edge [] node [above, near start] {\scriptsize{$\frac{1}{2}$}} (s12)
      (s11) edge [] (s111)
      (s12) edge [] (s121)
      (s111) edge [] (s1111)
      (s121) edge [] (s1211)
      (s1211) edge [] (s1211b)
      (s1211b) edge [] (s1211bb)
      (s1211bb) edge [] (undesired)
      (s1111) edge [] node [above, near start] {\scriptsize{$\frac{1}{2}$}} (s11111)
      (s1111) edge [] node [above, near start] {\scriptsize{$\frac{1}{2}$}} (s11112)
      (s11111) edge [] node [right, near start] {} (s111111)	
      (s11112) edge [] node [right, near start] {} (s111121)	
      (s111111) edge [] node [right, near start] {} (s1111111)	
      (s111121) edge [] node [right, near start] {} (s1111211)	
      (s1111211) edge [] node [right, near start] {} (s1111211b)
      (s1111211b) edge [] node [right, near start] {} (s1111211bb)	
      (s1111211bb) edge [] node [right, near start] {} (s11112111)
      (s11112111) edge [] node [right, near start] {} (sink)	
      (undesired) edge [] (sink)
      (sink) edge [loop right] (sink)
;
\end{tikzpicture}}	
\end{center}
\caption{Partially unrolled operational semantics for program $P$}	
\label{fig:better_operational_example}
\end{figure}
\begin{example}

%
%
Reconsider Example~\ref{ex:simple}, where we set for readability $P_1= \PChoice{\Ass{x}{x + 1}}{0.5}{\Ass{c}{1}}$, $P_2= \Observe (\textit{``$x$ is odd"})$, $P_3= \{\Ass{x}{x+1}\}$, and $P_4=\{\Ass{c}{1}\}$. 
A part of the operational MDP $\Rmdp{}{f}{P}$ for an arbitrary initial variable valuation $\sigma_I$ and post--expectation $x$ is depicted in Figure~\ref{fig:better_operational_example}.\footnote{We have tacitly overloaded the variable name $x$ to an expectation here for readability. More formally, by the ``expectation $x$" we actually mean the expectation $\lambda \sigma.~ \sigma(x)$.}
Note that this MDP is an MC, as $P$ contains no nondeterministic choices. The MDP has been unrolled until the second loop iteration, \ie, at state $\langle P,\,\sigma_I[x/2]\rangle$, the unrolling could be continued. The only terminating state is $\langle \term,\,\sigma_I[x/1,c/1]\rangle$. 
As our post-expectation is the value of variable $x$, we assign this value to terminating states, \ie, reward $\boxed{1}$ at state $\langle \term,\,\sigma_I[x/1,c/1]\rangle$, where $x$ has been assigned $1$. At state $\langle P,\,\sigma_I[c/1]\rangle$, the loop condition is violated as is the subsequent observation because of $x$ being assigned an even number.
\hfill$\triangle$
\end{example}

\section{Bounded Model Checking for Probabilistic Programs}
\label{sec:bmc}
In this section we describe our approach to model checking probabilistic programs. The key idea is that satisfaction or violation of certain properties for a program can be shown by means of a \emph{finite unrolling} of the program. Therefore, we introduce the notion of a partial operational semantics of a program, which we exploit to apply standard model checking to prove or disprove properties.

\noindent First, we state the correspondence between the satisfaction of a property for a \cpGCL-program $P$ and for its operational semantics, the MDP $\Rmdp{}{f}{P}$.
Intuitively, a program satisfies a property if and only if the property is satisfied on the operational semantics of the program.

\begin{definition}[Satisfaction of Properties]\label{def:property_satisfaction}
	Given a \cpGCL program $P$ and a (conditional) reachability or expected reward property $\varphi$. We define
	\begin{align*}
		P\models\varphi \quad \text{ iff } \quad \Rmdp{}{f}{P}\models\varphi~.
	\end{align*}
\end{definition}
This correspondence on the level of a denotational semantics for \cpGCL has been discussed extensively in~\cite{jansen-et-al-mfps-2015}. Note that there only schedulers which minimize expected rewards were considered. Here, we also need maximal schedulers as we are considering both upper and lower bounds on expected rewards and probabilities. Note that satisfaction of properties is solely based on the operational semantics and induced maximal or minimal probabilities or expected rewards.

We now introduce the notion of a partial operational MDP for a \cpGCL--program $P$, which is a finite approximation of the full operational MDP of $P$. Intuitively, this amounts to the successive application of SOS rules given in Figure~\ref{fig:sos-rules}, while not all possible rules have been applied yet.
\begin{definition}[Partial Operational Semantics]\label{def:partial_mdp}
	A \emph{partial operational semantics} for a \cpGCL--program $P$ is a sub-MDP $\Rmdp{}{f}{P}'=(S', \langle P,\,\sigma_I\rangle,\, \Act,\, \pdtmc')$ of the operational semantics for $P$ (denoted $\Rmdp{}{f}{P}'\subseteq\Rmdp{}{f}{P}$) with $S'\subseteq S$.
Let $S_{\textit{exp}}= S' \setminus \big\{\langle Q,\sigma\rangle\in S' ~\big|~ Q\neq\term,~ \exists\, s \in S \setminus S'~\exists\,  \alpha \in \Act\colon \pdtmc\big(\langle Q,\, \sigma\rangle,\, \alpha,\, s\big) > 0 \big\}$ be the \emph{set of expandable states}. Then the transition probability function $\pdtmc'$ is for $s,s'\in S'$ and $\act\in\Act$ given by
		\begin{align*}
			\pmdp'(s,\act,s') ~=~
			\begin{cases}
				1, & \textit{if }s=s' \textit{ for } s,s'\in S_{\textit{exp}},\\
				\pmdp(s,\act,s'), & \textit{otherwise .}
			\end{cases}
		\end{align*}
\end{definition}
Intuitively, the set of non--terminating \emph{expandable states} describes the states where there are still SOS rules applicable. 
Using this definition, the only transitions leaving expandable states are self-loops, enabling to have a well-defined probability measure on partial operational semantics. We will use this for our method, which is based on the fact that both (conditional) reachability probabilities and expected rewards for certain properties will always monotonically increase for further unrollings of a program and the respective partial operational semantics. This is discussed in what follows.

\subsection{Growing Expectations}
As mentioned before, we are interested in the probability of termination or the expected values of expectations (i.e.\ random variables ranging over program states) after successful termination of the program.
This is measured on the operational MDP by the set of paths \emph{reaching $\sink$ from the initial state conditioned on not reaching $\undesired$}~\cite{jansen-et-al-mfps-2015}. In detail, we have to compute the conditional expected value of post--expectation $f$ after successful termination of program $P$, given that no observation was violated along the computation. For nondeterministic programs, we have to compute this value either under a minimizing or maximizing scheduler (depending on the given property). We focus our presentation on expected rewards and minimizing schedulers, but all concepts are analogous for the other cases. For $\Rmdp{}{f}{P}$ we have 
\begin{align*}
	\inf_{\sched\in\Sched^{\Rmdp{}{f}{P}}}\cexpRew{\Rmdp{}{f}{P}^\sched}{\sink}{\undesired}~.
\end{align*} 
Recall that $\Rmdp{}{f}{P}^\sched$ is the induced MC under scheduler $\sched\in\Sched^{\Rmdp{}{f}{P}}$ as in Definition~\ref{def:induced_dtmc}. Recall also that for $\neg\finally\undesired$ all paths not eventually reaching \undesired either diverge (collecting reward 0) or pass by a \exit--state and reach \sink.
More importantly, all paths that \emph{do} eventually reach \undesired also collect reward 0. 
Thus:
%
%
\begin{align*}
	\inf_{\sched\in\Sched^{\Rmdp{}{f}{P}}}& \cexpRew{\Rmdp{}{f}{P}^\sched}{\sink}{\undesired}\\
	~=~ \inf_{\sched\in\Sched^{\Rmdp{}{f}{P}}} & \frac{\expRew{\Rmdp{}{f}{P}^\sched}{\sink \cap \neg\finally\undesired}}{\Pr^{{\Rmdp {} {f} P}^\sched}({\neg\lozenge\bad})}\\
	~=~ \inf_{\sched\in\Sched^{\Rmdp {} {f} P}} & \frac{\expRew{\Rmdp{}{f}{P}^\sched}{\sink}}{\Pr^{{\Rmdp {} {f} P}^\sched}({\neg\lozenge\bad})}~.
	\intertext{Finally, observe that the probability of not reaching \undesired is one minus the probability of reaching \undesired, which gives us:}
	~=~ \inf_{\sched\in\Sched^{\Rmdp {} {f} P}} & \frac{\expRew{\Rmdp{}{f}{P}^\sched}{\sink}}{1 - \Pr^{{\Rmdp {} {f} P}^\sched}({\finally\bad})}~.\tag{$\dagger$}
\end{align*}
Regarding the quotient minimization we assume ``$\frac 0 0 < 0$" as we see $\frac 0 0$---being undefined---to be less favorable than $0$.
For programs without nondeterminism this view agrees with a weakest--precondition--style semantics for probabilistic programs with conditioning~\cite{jansen-et-al-mfps-2015}.

It was shown in~\cite{KaKa2015} that \emph{all strict lower bounds} for $\expRew{\Rmdp{}{f}{P}^\sched}{\sink}$ are in principle computably enumerable in a monotonically non--decreasing fashion.
One way to do so, is to allow for the program to be executed for an increasing number of $k$ steps, and collect the expected rewards of all execution traces that have lead to termination within $k$ computation steps.
This corresponds naturally to constructing a partial operational semantics  $\Rmdp{}{f}{P}'\subseteq\Rmdp{}{f}{P}$ as in Definition~\ref{def:partial_mdp} and computing minimal expected rewards on $\Rmdp{}{f}{P}'$.

Analogously, it is of course also possible to monotonically enumerate all strict lower bounds of $\Pr^{{\Rmdp {} {f} P}^\sched}({\finally\bad})$, since---again---we need to just collect the probability mass of all traces that have led to \undesired within $k$ computation steps.
Since probabilities are quantities bounded between 0 and 1, a lower bound for $\Pr^{{\Rmdp{}{f}{P}}^\sched}({\finally\bad})$ is an upper bound for $1 - \Pr^{{\Rmdp {} {f} P}^\sched}({\finally\bad})$.

Put together, a lower bound for $\expRew{\Rmdp{}{f}{P}^\sched}{\sink}$ and a lower bound for $\Pr^{{\Rmdp {} {f} P}^\sched}({\finally\bad})$ yields a lower bound for ($\dagger$).
We are thus able to enumerate all lower bounds of $\cexpRew{\Rmdp{}{f}{P}^\sched}{\sink}{\undesired}$ by inspection of a finite sub--MDP of $\Rmdp{}{f}{P}$.
Formally, we have:
\begin{theorem}\label{theo:correctness}
	For a \cpGCL program $P$, post--expectation $f$, and a partial operational MDP $\Rmdp{}{f}{P}'\subseteq \Rmdp{}{f}{P}$ it holds that	
\begin{align*}
		&\inf_{\sched\in\Sched^{\Rmdp{}{f}{P}'}} \cexpRew{\Rmdp{}{f}{P}'^\sched}{\sink}{\undesired}\\
		&\qquad\qquad \leq~ \inf_{\sched\in\Sched^{\Rmdp{}{f}{P}}} \cexpRew{\Rmdp{}{f}{P}^\sched}{\sink}{\undesired}~.
	\end{align*}
\end{theorem}

%

\subsection{Model Checking}
Using Theorem~\ref{theo:correctness}, we transfer satisfaction or violation of certain properties from a partial operational semantics $\Rmdp{}{f}{P}'\subseteq\Rmdp{}{f}{P}$ to the full semantics of the program. 
For an upper bounded conditional expected reward property $\reachPropSymbol = \cexpRewPropkT $ where $T,U\in\State$ we exploit that 
	\begin{align}
		\Rmdp{}{f}{P}'\not\models\reachPropSymbol\quad \implies \quad P\not\models\reachPropSymbol\ .\label{eq:propviolation}
	\end{align}
	That means, if we can prove the violation of $\reachPropSymbol$ on the MDP induced by a finite unrolling of the program, it will hold for all further unrollings, too.
This is because all rewards and probabilities are positive and thus further unrolling can only increase the accumulated reward and/or probability mass.
	
	Dually, for a lower bounded conditional expected reward property $\psi = \rew_{\geq\lambda}(\finally T \mid \finally U)$ we use the following property:
	\begin{align}
		\Rmdp{}{f}{P}'\models\psi\quad \implies \quad P\models\reachPropSymbol\ .\label{eq:propsatisfaction}
	\end{align}
	The preconditions of Implication~(\ref{eq:propviolation}) and Implication~(\ref{eq:propsatisfaction}) can be checked by probabilistic model checkers like \prism~\cite{KNP11}; this is analogous for conditional reachability properties. Let us illustrate this by means of an example.
\begin{example}
	As mentioned in Example~\ref{ex:simple}, we are interested in the \emph{probability of termination}. As outlined in Section~\ref{sec:operational}, this probability can be measured by 
		\begin{align*}
			\pr(\finally\sink\mid\neg\finally\undesired) ~=~\frac{\pr(\finally\sink\land\neg\finally\undesired)}{\pr(\finally\undesired)}\ .
		\end{align*}
		 We want this probability to be at least $\nicefrac1 2$, \ie, $\varphi=\p_{\geq 0.5}(\lozenge \sink \mid \neg\lozenge\undesired)$. 
		 Since for further unrollings of our partially unrolled MDP this probability never decreases, the property can already be verified on the partial MDP $\Rmdp{}{f}{P}'$ by
		 \begin{align*}
		 	\pr^{\Rmdp{}{f}{P}'}(\finally\sink\mid\neg\finally\undesired) ~=~ \frac{\nicefrac{1}{4}}{\nicefrac{1}{2}} ~=~ \frac{1}{2}\ ,
		 \end{align*}
		 where $\Rmdp{}{f}{P}'$ is the sub-MDP from Figure \ref{fig:better_operational_example}. This finite sub-MDP $\Rmdp{}{f}{P}'$ is therefore a witness of $\Rmdp{}{f}{P}\models\varphi$. 
\hfill$\triangle$
\end{example}
	Algorithmically, this technique relies on suitable heuristics regarding the size of the considered partial MDPs. Basically, in each step $k$ states are expanded and the corresponding MDP is model checked, until either the property can be shown to be satisfied or violated, or no more states are expandable. In addition, heuristics based on shortest path searching algorithms can be employed to favor expandable states that so far induce high probabilities. 
	
	Note that this method is a \emph{semi-algorithm} when the model checking problems stated in Implications~\eqref{eq:propviolation} and~\eqref{eq:propsatisfaction} are considering strict bounds, i.e.\ $<\kappa$ and $> \kappa$.
	 It is then guaranteed that the given bounds are finally exceed.
	
	Consider now the case where we want to show \emph{satisfaction} of $\reachPropSymbol = \cexpRewPropkT$, \ie, $\Rmdp{}{f}{P}'\models\reachPropSymbol\ \Rightarrow\ P\models\reachPropSymbol$. As the conditional expected reward will monotonically increase as long as the partial MDP is expandable, the implication is only true if there are no more expandable states, \ie, the model is fully expanded. This is analogous for the violation of upper bounded properties. Note that many practical examples actually induce finite operational MDPs which enables to build the full model and perform model checking. 
	
	It remains to discuss how this approach can be utilized for parameter synthesis as explained in Section~\ref{sec:probmodels}. For a partial operational pMDP $\Rmdp{}{f}{P}'$ and a property $\reachPropSymbol = \cexpRewPropkT$ we use tools like \prophesy~\cite{dehnert-et-al-cav-2015} to determine for which parameter valuations $\reachPropSymbol$ is violated. For each valuation  $u$ with $\Rmdp{}{f}{P}'[u]\not\models\reachPropSymbol$ it holds that $\Rmdp{}{f}{P}[u]\not\models\reachPropSymbol$; each parameter valuation violating a property on a partial pMDP also violates it on the fully expanded MDP.
	
	%
%

%
\section{Evaluation}
\label{sec:evaluation} 
\paragraph{Experimental Setup.}
We implemented and evaluated the bounded model checking method in C++. For the model checking functionality, we use the stochastic model checker \storm, developed at RWTH Aachen University, and \prophesy~\cite{dehnert-et-al-atva-2014} for parameter synthesis. 

We consider five different, well-known benchmark programs, three of which are based on models from the \prism benchmark suite~\cite{KNP11} and others taken from other literature (see Appendix~\ref{app:models} for some examples). We give the running times of our prototype on several instances of these models. Since there is --- to the best of our knowledge --- no other tool that can analyze \cpGCL~programs in a purely automated fashion, we cannot meaningfully compare these figures to other tools. As our technique is restricted to establishing that lower bounds on reachability probabilities and the expectations of program variables, respectively, exceed a threshold $\lambda$, we need to fix $\lambda$ for each experiment. For all our experiments, we chose $\lambda$ to be 90\% of the actual value for the corresponding query and choose to expand $10^6$ states of the partial operational semantics of a program between each model checking run.

 We ran the experiments on an HP BL685C G7 machine with 48 cores clocked with 2.0GHz each and 192GB of RAM while each experiment only runs in a single thread with a time--out of one hour.
   We ran the following benchmarks\footnote{All input programs and log files of the experiments can be downloaded at \href{https://moves.rwth-aachen.de/wp-content/uploads/conference\_material/pgcl\_atva16.tar.gz}{moves.rwth-aachen.de/wp-content/uploads/conference\_material/pgcl\_atva16.tar.gz}}:
\paragraph{Crowds Protocol~\textnormal{\cite{RR98}}.}
This protocol aims at anonymizing the sender of $R$ messages by routing them probabilistically through a crowd of $N$ hosts. Some of these hosts, however, are corrupt and try to determine the real sender by observing the host that most recently forwarded a message. For this model, we are interested in \begin{enumerate*}[label={(\alph*)}] \item the probability that the real sender is observed more than $R/10$ times, and \item the expected number of times that the real sender is observed. \end{enumerate*}

We also consider a variant (crowds-obs) of the model in which an observe statement ensures that after all messages have been delivered, hosts different from the real sender have been observed at least $R/4$ times. Unlike the model from the PRISM website, our model abstracts from the concrete identity of hosts different from the sender, since they are irrelevant for  properties of interest.
\paragraph{Herman Protocol.}
In this protocol~\cite{herman-self-stabilization}, $N$ hosts form a token-passing ring and try to steer the system into a stable state. We consider the probability that the system eventually reaches such a state in two variants of this model where the initial state is either chosen probabilistically or nondeterministically.
\paragraph{Robot.}
The robot case-study is loosely based on a similar model from the PRISM benchmark suite. It models a robot that navigates through a bounded area of an unbounded grid. Doing so, the robot can be blocked by a janitor that is moving probabilistically across the whole grid. The property of interest is the probability that the robot will eventually reach its final destination.
\paragraph{Predator.}
This model is due to Lotka and Volterra~\cite[p. 127]{lotkavolterra}. A predator and a prey population evolve with mutual dependency on each other's numbers. Following some basic biology principles, both populations undergo periodic fluctuations. We are interested in \begin{enumerate*}[label={(\alph*)}] \item the probability of one of the species going extinct, and \item the expected size of the prey population after one species has gone extinct. \end{enumerate*}
\paragraph{Coupon Collector.}
This is a famous example\footnote{\href{https://en.wikipedia.org/wiki/Coupon\_collector\%27s\_problem}{https://en.wikipedia.org/wiki/Coupon\_collector\%27s\_problem}} from textbooks on randomized algorithms \cite{erdoes-coupon-collector}. A collector's goal is to collect all of $N$ distinct coupons. In every round, the collector draws three new coupons chosen uniformly at random out of the $N$ coupons. We consider \begin{enumerate*}[label={(\alph*)}] \item the probability that the collector possesses all coupons after $N$ rounds, and \item the expected number of rounds the collector needs until he has all the coupons \end{enumerate*} as properties of interest. Furthermore, we consider two slight variants: in the first one (coupon-obs), an observe statement ensures that the three drawn coupons are all different and in the second one (coupon-classic), the collector may only draw one coupon in each round.
%
{
  \setlength{\tabcolsep}{4pt}
  \begin{table}[t]
    \caption{Benchmark results for probability queries.}
    \centering
    \begin{tabular}{ccrrrrrrrr}
      \hline
      program & instance & \#states & \#trans. & full? & $\lambda$ & result & actual & time \\
      
      \hline
      \hline

      \multirow{3}{*}{crowds} & (100,60) & $877370$ & $1104290$ & yes &  $0.29$ & $0.33$ & $0.33$ & $109$ \\
      
      & (100,80) & $10^6$ & $1258755$ & no & $0.30$ & $0.33$ & $0.33$ & $131$ \\
      
      & (100,100) & $2 \cdot 10^6$ & $2518395$ & no & $0.30$ & $0.33$ & $0.33$ & $354$ \\
      
      \hline

      \multirow{3}{*}{crowds-obs} & (100,60) & $878405$ & $1105325$ & yes & $0.23$ & $0.26$ & $0.26$ & $126$ \\
      
      & (100,80) & $10^6$ & $1258718$ & no & $0.23$ & $0.25$ & $0.24$ & $170$ \\
      
      & (100,100) & $3 \cdot 10^6$ & $3778192$ & no & $0.23$ & $0.26$ & $0.26$ & $890$ \\
      
      \hline
      
      \multirow{2}{*}{herman} & (17) & $10^6$ & $1136612$ & no & $0.9$ & $0.99$ & $1$ & $91$ \\

      & (21) & $10^6$ & $1222530$ & no & $0.9$ & $0.99$ & $1$ & $142$ \\

      \hline

      \multirow{2}{*}{herman-nd} & (13) & $1005945$ & $1112188$ & yes & $0.9$ & $1$ & $1$ & $551$ \\

      & (17) & $-$ & $-$ & no & $0.9$ & $0$ & $1$ & TO \\

      \hline

      robot & - & $181595$ & $234320$ & yes & $0.9$ & $1$ & $1$ & $24$ \\

      \hline
      
      predator & - & $10^6$ & $1234854$ & no & $0.9$ & $0.98$ & $1$ & $116$ \\

      \hline

      \multirow{3}{*}{coupon} & (5) & $10^6$ & $1589528$ & no & $0.75$ & $0.83$ & $0.83$ & $11$ \\

      & (7) & $2 \cdot 10^6$ & $3635966$ & no & $0.67$ & $0.72$ & $0.74$ & $440$ \\

      & (10) & $-$ & $-$ & no & $0.57$ & $0$ & $0.63$ & TO \\

      \hline

      \multirow{3}{*}{coupon-obs} & (5) & $10^6$ & $1750932$ & no & $0.85$ & $0.99$ & $0.99$ & $11$ \\

      & (7) & $10^6$ & $1901206$ & no & $0.88$ & $0.91$ & $0.98$ & $15$ \\

      & (10) & $-$ & $-$ & no & $0.85$ & $0$ & $0.95$ & TO \\

      \hline

      \multirow{3}{*}{coupon-classic} & (5) & $10^6$ & $1356463$ & no & 3.4e-3 & 3.8e-3 & 3.8e-3 & $9$ \\

      & (7) & $10^6$ & $1428286$ & no & 5.5e-4 & 6.1e-4 & 6.1e-4 & $9$ \\

      & (10) & $-$ & $-$ & no & 3.3e-5 & $0$ & 3.6e-5 & TO \\

      \hline

    \end{tabular}
    \vspace{0.2cm}
    \label{table:probqueries}
  \end{table}
}

Table \ref{table:probqueries} shows the results for the probability queries. For each model instance, we give the number of explored states and transitions and whether or not the model was fully expanded. Note that the state number is a multiple of $10^6$ in case the model was not fully explored, because our prototype always expands $10^6$ states before it does the next model checking call. The next three columns show the probability bound ($\lambda$), the result that the tool could achieve as well as the actual answer to the query on the full (potentially infinite) model. Due to space constraints, we rounded these figures to two significant digits. We report on the time in seconds that the prototype took to establish the result (TO = 3600 sec.).

We observe that for most examples it suffices to perform few unfolding steps to achieve more than 90\% of the actual probability. For example, for the largest crowds-obs program, $3 \cdot 10^6$ states are expanded, meaning that three unfolding steps were performed. Answering queries on programs including an observe statement can be costlier (crowds vs. crowds-obs), but does not need to be (coupon vs. coupon-obs). In the latter case, the observe statement prunes some paths early that were not promising to begin with, whereas in the former case, the observe statement only happens at the very end, which intuitively makes it harder for the search to find target states. We are able to obtain non-trivial lower bounds for all but two case studies. For herman-nd, not all of the (nondeterministically chosen) initial states were explored, because our exploration order currently  does not favour states that influence the obtained result the most. Similarly, for the largest coupon collector examples, the time limit did not allow for finding one target state. Again, an exploration heuristic that is more directed towards these could potentially improve performance drastically.

{
  \setlength{\tabcolsep}{4pt}
  \begin{table}[t]
  	\caption{Benchmark results for expectation queries.}
    \centering
    \begin{tabular}{ccrrrrrrrr}
      \hline
      program & instance & \#states & \#trans. & full? & result & actual & time \\
      
      \hline
      \hline

      \multirow{3}{*}{crowds} & (100,60) & $877370$ & $1104290$ & yes & $5.61$ & $5.61$ & $125$ \\
      
      & (100,80) & $10^6$ & $1258605$ & no & $7.27$ & $7.47$ & $176$ \\
      
      & (100,100) & $2 \cdot 10^6$ & $2518270$ & no & $9.22$ & $9.34$ & $383$ \\
      
      \hline

      \multirow{3}{*}{crowds-obs} & (100,60) & $878405$ & $1105325$ & yes & $5.18$ & $5.18$ & $134$ \\
      
      & (100,80) & $10^6$ & $1258569$ & no & $6.42$ & $6.98$ & $206$ \\
      
      & (100,100) & $2 \cdot 10^6$ & $2518220$ & no & $8.39$ & $8.79$ & $462$ \\
      
      \hline
            
      predator & $-$ & $3 \cdot 10^6$ & $3716578$ & no & $99.14$ & ? & $369$ \\

      \hline

      \multirow{3}{*}{coupon} & (5) & $10^6$ & $1589528$ & no & $4.13$ & $4.13$ & $15$ \\

      & (7) & $3 \cdot 10^6$ & $5379492$ & no & $5.86$ & $6.38$ & $46$ \\

      & (10) & $-$ & $-$ & no & $0$ & $10.1$ & TO \\

      \hline

      \multirow{3}{*}{coupon-obs} & (5) & $10^6$ & $1750932$ & no & $2.57$ & $2.57$ & $13$ \\

      & (7) & $2 \cdot 10^6$ & $3752912$ & no & $4.22$ & $4.23$ & $30$ \\

      & (10) & $-$ & $-$ & no & $0$ & $6.96$ & TO \\

      \hline

      \multirow{3}{*}{coupon-classic} & (5) & $10^6$ & $1356463$ & no & $11.41$ & $11.42$ & $15$ \\

      & (7) & $10^6$ & $1393360$ & no & $18.15$ & $18.15$ & $21$ \\

      & (10) & $-$ & $-$ & no & $0$ & $29.29$ & TO \\

      \hline
    \end{tabular}
    \vspace{0.2cm}
    \label{table:expqueries}
  \end{table}
}
Table \ref{table:expqueries} shows the results for computing the expected value of program variables at terminating states. For technical reasons, our prototype currently cannot perform more than one unfolding step for this type of query. To achieve meaningful results, we therefore vary the number of explored states until 90\% of the actual result is achieved. Note that for the predator program, the actual value for the query is not known to us, so we report on the value at which the result only grows very slowly. The results are similar to the probability case in that most often a low number of states suffices to show meaningful lower bounds. Unfortunately --- as before --- we can only prove a trivial lower bound for the largest coupon collector examples.

\begin{figure}[t]
  \subfigure[coupon-obs (5)]{
    \includegraphics[scale=0.43]{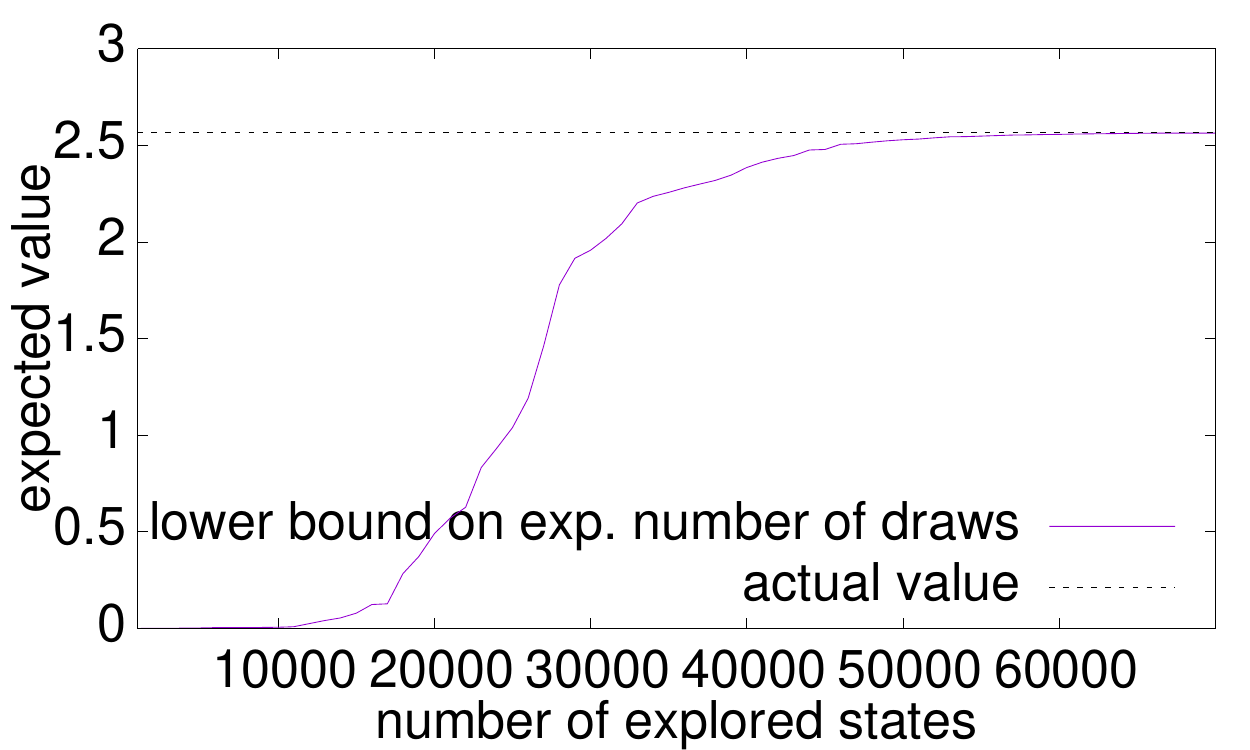}
  }
  \hfill
  \subfigure[predator]{
    \includegraphics[scale=0.43]{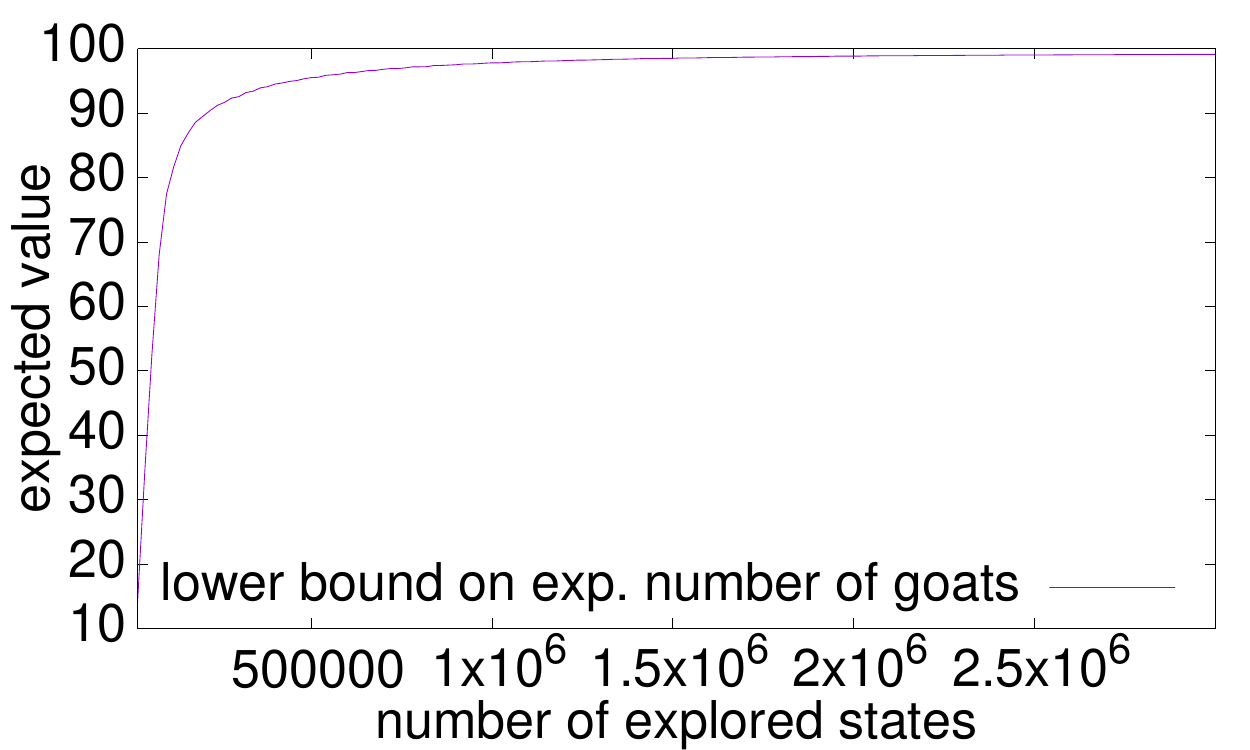}
  }
  \caption{The obtained values approach the actual value from below.}
  \label{fig:lower_bound_approaching_actual_value}
\end{figure}
Figure~\ref{fig:lower_bound_approaching_actual_value} illustrates how the obtained lower bounds approach the actual expected value with increasing number of explored states for two case studies. For example, in the left picture one can observe that exploring 60000 states is enough to obtain a very precise lower bound on the expected number of rounds the collector needs to gather all five coupons, as indicated by the dashed line.

\begin{figure}[t]

  \subfigure[after 9 iterations]{
    \scalebox{0.8}{\includegraphics[scale=0.44]{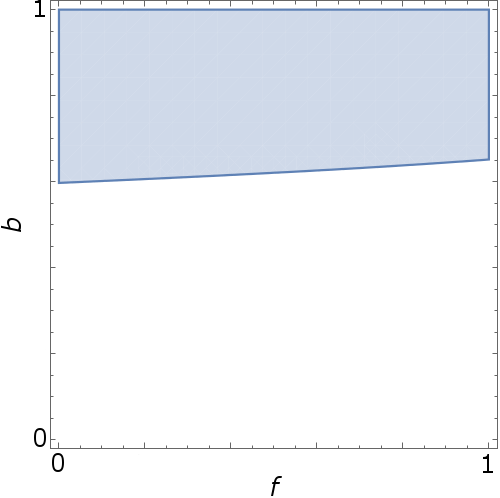}}
  }
  \hfill
  \subfigure[after 13 iterations]{
    \scalebox{0.8}{\includegraphics[scale=0.44]{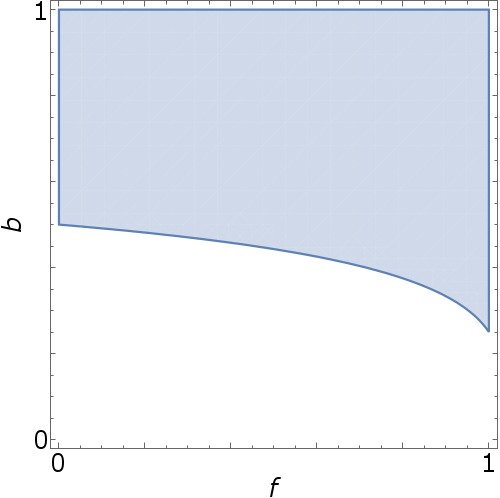}}
  }
  \caption{Analyzing parametric models yields violating parameter instances.}  
  \label{fig:lower_bounds_parametric_model}
\end{figure}
Finally, we analyze a parametric version of the crowds model that uses the parameters $f$ and $b$ to leave the probabilities \begin{enumerate*}[label={(\roman*)}] \item for a crowd member to be corrupt ($b$) and \item of forwarding (instead of delivering) a message ($f$) \end{enumerate*} unspecified. In each iteration of our algorithm, we obtain a rational function describing a lower bound on the actual probability of observing the real sender of the message more than once \emph{for each parameter valuation}. Figure \ref{fig:lower_bounds_parametric_model} shows the regions of the parameter space in which the protocol was determined to be unsafe (after iterations $9$ and $13$, respectively) in the sense that the probability to identify the real sender exceeds $\frac{1}{2}$. Since the results obtained over different iterations are monotonically increasing, we can conclude that all parameter valuations that were proved to be unsafe in some iteration are in fact unsafe in the full model. This in turn means that the blue area in Figure \ref{fig:lower_bounds_parametric_model} grows in each iteration.

\section{Conclusion and Future Work}
\label{sec:conclusion}
We presented a direct verification method for probabilistic programs employing probabilistic model checking. 
We conjecture that the basic idea would smoothly translate to reasoning about recursive probabilistic programs~\cite{recursion}.
In the future we are interested in how loop invariants~\cite{DBLP:conf/qest/GretzKM13} can be utilized to devise complete model checking procedures preventing possibly infinite loop unrollings. 
This is especially interesting for reasoning about covariances \cite{covariances}, where a mixture of invariant--reasoning and successively constructing the operational MC would yield sound over- and underapproximations of covariances.
To extend the gain for the user, we will combine this approach with methods for counterexamples~\cite{DBLP:conf/sfm/AbrahamBDJKW14}, which can be given in terms of the programming language~\cite{wimmer-et-al-lmcs-2015,dehnert-et-al-atva-2014}. Moreover, it seem promising to investigate how approaches to automatically \emph{repair} a probabilistic model towards satisfaction of properties~\cite{DBLP:conf/tacas/BartocciGKRS11,DBLP:conf/nfm/PathakAJTK15} can be transferred to programs.

\bibliographystyle{splncs}
\bibliography{bibliography}

\clearpage
\appendix
\section{Models}
\label{app:models}
\lstset{tabsize=2}
\subsection{coupon-obs (5)}
\begin{lstlisting}
int coup0 := 0;
int coup1 := 0;
int coup2 := 0;
int coup3 := 0;
int coup4 := 0;

int draw1 := 0;
int draw2 := 0;
int draw3 := 0;

int numberDraws := 0;

while (!(coup0 = 1) | !(coup1 = 1) | !(coup2 = 1) | !(coup3 = 1) | !(coup4 = 1)) {
	draw1 := unif(0,4);
	draw2 := unif(0,4);
	draw3 := unif(0,4);
	numberDraws := numberDraws + 1;
		
	observe (draw1 != draw2 & draw1 != draw3 & draw2 != draw3);

	if(draw1 = 0 | draw2 = 0 | draw3 = 0) {
		coup0 := 1;
	}
	if(draw1 = 1 | draw2 = 1 | draw3 = 1) {
		coup1 := 1;
	}
	if(draw1 = 2 | draw2 = 2 | draw3 = 2) {
		coup2 := 1;
	}
	if (draw1 = 3 | draw2 = 3 | draw3 = 3) {
		coup3 := 1;
	}
	if (draw1 = 4 | draw2 = 4 | draw3 = 4) {
		coup4 := 1;
	}
}
\end{lstlisting}

\clearpage
\subsection{coupon (5)}
\begin{lstlisting}
int coup0 := 0;
int coup1 := 0;
int coup2 := 0;
int coup3 := 0;
int coup4 := 0;

int draw1 := 0;
int draw2 := 0;
int draw3 := 0;

int numberDraws := 0;

while (!(coup0 = 1) | !(coup1 = 1) | !(coup2 = 1) | !(coup3 = 1) | !(coup4 = 1)) {
	draw1 := unif(0,4);
	draw2 := unif(0,4);
	draw3 := unif(0,4);
	numberDraws := numberDraws + 1;
	
	if(draw1 = 0 | draw2 = 0 | draw3 = 0) {
		coup0 := 1;
	}
	if(draw1 = 1 | draw2 = 1 | draw3 = 1) {
		coup1 := 1;
	}
	if(draw1 = 2 | draw2 = 2 | draw3 = 2) {
		coup2 := 1;
	}
	if (draw1 = 3 | draw2 = 3 | draw3 = 3) {
		coup3 := 1;
	}
	if (draw1 = 4 | draw2 = 4 | draw3 = 4) {
		coup4 := 1;
	}
}
\end{lstlisting}

\clearpage
\subsection{crowds-obs (100, 60)}
\begin{lstlisting}
int delivered := 0;
int lastSender := 0;
int remainingRuns := 60;
int observeSender := 0;
int observeOther := 0;

while(remainingRuns > 0) {
	while(delivered = 0) {
		{
			if(lastSender = 0) {
				observeSender := observeSender + 1;
			} else {
				observeOther := observeOther + 1;
			}
			lastSender := 0;
			delivered := 1;
		} [0.091] {
			{
				{ lastSender:=0; } [1/100] { lastSender := 1; }
			}
			[0.8]
			{
				lastSender := 0;
				// When not forwarding, the message is delivered here
				delivered := 1;
			}
		}
	}
	// Set up new run.
	delivered := 0;
	remainingRuns := remainingRuns - 1;
}
observe(observeOther > 15);
\end{lstlisting}

\clearpage
\subsection{crowds (100, 60)}
\begin{lstlisting}
int delivered := 0;
int lastSender := 0;
int remainingRuns := 60;
int observeSender := 0;
int observeOther := 0;

while(remainingRuns > 0) {
	while(delivered = 0) {
		{
			if(lastSender = 0) {
				observeSender := observeSender + 1;
			} else {
				observeOther := observeOther + 1;
			}
			lastSender := 0;
			delivered := 1;
		} [0.091] {
			{
				{ lastSender:=0; } [1/100] { lastSender := 1; }
			}
			[0.8]
			{
				lastSender := 0;
				// When not forwarding, the message is delivered here
				delivered := 1;
			}
		}
	}
	// Set up new run.
	delivered := 0;
	remainingRuns := remainingRuns - 1;
}
\end{lstlisting}

\clearpage
\subsection{crowds (100, 60) parametric}
This program is parametric with the parameters $f$ (probability of forwarding the message) and $b$ (probability that a crowd member is bad).
\begin{lstlisting}
int delivered := 0;
int lastSender := 0;
int remainingRuns := 60;
int observeSender := 0;
int observeOther := 0;

while(remainingRuns > 0) {
	while(delivered = 0) {
		{
			if(lastSender = 0) {
				observeSender := observeSender + 1;
			} else {
				observeOther := observeOther + 1;
			}
			lastSender := 0;
			delivered := 1;
		} [b] {
			{
				{ lastSender:=0; } [1/100] { lastSender := 1; }
			}
			[f]
			{
				lastSender := 0;
				// When not forwarding, the message is delivered here
				delivered := 1;
			}
		}
	}
	// Set up new run.
	delivered := 0;
	remainingRuns := remainingRuns - 1;
}
\end{lstlisting}

\end{document}